\documentclass[aps,prl,twocolumn,showpacs,groupedaddress,superscriptaddress]{revtex4-1}
\usepackage{amssymb}
\usepackage{eurosym}
\usepackage{dcolumn}
\usepackage{bm}
\usepackage{graphicx}
\usepackage{color}
\usepackage{amsmath}
\usepackage{amsfonts}

\begin{document}
\title{Magnetization process of atacamite: a case of weakly coupled $S = 1/2$ sawtooth chains}
\author{L. Heinze}
\affiliation{Institut f\"ur Physik der Kondensierten Materie, TU Braunschweig, D-38106 Braunschweig, Germany}
\author{H. O. Jeschke}
\affiliation{Research Institute for Interdisciplinary Science, Okayama University, Okayama 700-8530, Japan}
\author{I. I. Mazin}
\affiliation{Department of Physics and Astronomy, George Mason University, Fairfax, Virginia 22030, USA}
\affiliation{Quantum Science and Engineering Center, George Mason University, Fairfax, Virginia 22030, USA}
\author{A. Metavitsiadis}
\affiliation{Institut f\"ur Theoretische Physik, TU Braunschweig, D-38106 Braunschweig, Germany}
\author{M. Reehuis}
\affiliation{Helmholtz-Zentrum Berlin f\"ur Materialien und Energie GmbH, D-14109 Berlin, Germany}
\author{R. Feyerherm}
\affiliation{Helmholtz-Zentrum Berlin f\"ur Materialien und Energie GmbH, D-14109 Berlin, Germany}
\author{J.-U. Hoffmann}
\affiliation{Helmholtz-Zentrum Berlin f\"ur Materialien und Energie GmbH, D-14109 Berlin, Germany}
\author{M. Bartkowiak}
\affiliation{Helmholtz-Zentrum Berlin f\"ur Materialien und Energie GmbH, D-14109 Berlin, Germany}
\author{O. Prokhnenko}
\affiliation{Helmholtz-Zentrum Berlin f\"ur Materialien und Energie GmbH, D-14109 Berlin, Germany}
\author{A. U. B. Wolter}
\affiliation{Institute for Solid State and Materials Research, Leibniz IFW Dresden, D-01069 Dresden, Germany}
\author{X. Ding}
\affiliation{National High Magnetic Field Laboratory, Los Alamos National Laboratory, Los Alamos, New Mexico 87545, USA}
\author{V. S. Zapf}
\affiliation{National High Magnetic Field Laboratory, Los Alamos National Laboratory, Los Alamos, New Mexico 87545, USA}
\author{C. Corval\'an Moya}
\affiliation{National High Magnetic Field Laboratory, Los Alamos National Laboratory, Los Alamos, New Mexico 87545, USA}
\author{F. Weickert}
\affiliation{National High Magnetic Field Laboratory, Los Alamos National Laboratory, Los Alamos, New Mexico 87545, USA}
\author{M. Jaime}
\affiliation{National High Magnetic Field Laboratory, Los Alamos National Laboratory, Los Alamos, New Mexico 87545, USA}
\author{K. C. Rule}
\affiliation{Australian Centre for Neutron Scattering, Australian Nuclear Science and Technology Organisation, Lucas Heights, NSW 2234, Australia}
\author{D. Menzel}
\affiliation{Institut f\"ur Physik der Kondensierten Materie, TU Braunschweig, D-38106 Braunschweig, Germany}
\author{R. Valent\'i}
\affiliation{Institut f\"ur Theoretische Physik, Goethe-Universit\"at Frankfurt, D-60438 Frankfurt am Main, Germany}
\author{W. Brenig}
\affiliation{Institut f\"ur Theoretische Physik, TU Braunschweig, D-38106 Braunschweig, Germany}
\author{S. S\"{u}llow}
\affiliation{Institut f\"ur Physik der Kondensierten Materie, TU Braunschweig, D-38106 Braunschweig, Germany}
\date{\today}

\begin{abstract}
We present a combined experimental and theoretical study of the mineral atacamite Cu$_2$Cl(OH)$_3$. Density functional theory yields a Hamiltonian describing anisotropic sawtooth chains with weak 3D connections. Experimentally, we fully characterize the antiferromagnetically ordered state. Magnetic order shows a complex evolution with the magnetic field, while, starting at 31.5\,T, we observe a plateau-like magnetization at about $M_{\rm sat}/2$. Based on complementary theoretical approaches, we show that the latter is unrelated to the known magnetization plateau of a sawtooth chain. Instead, we provide evidence that the magnetization process in atacamite is a field-driven canting of a 3D network of weakly coupled sawtooth chains that form giant moments.
\end{abstract}
\maketitle

Frustrated low-dimensional quantum spin systems offer a unique opportunity to study complex quantum phases~\cite{lacroix2011,starykh2015,wosnitza2016,savary2017}. In the search for novel and exotic ground and field-induced states, such as spin liquids, magnetization plateaus or nematic phases, a multitude of models have been studied, including the kagome lattice, the diamond chain or the frustrated $J_{1}$-$J_{2}$ chain~\cite{harris1992,takano1996,tonegawa1987}. Experimental efforts to identify materials to test these theoretical concepts are exemplified by work on natural minerals such as herbertsmithite, azurite or linarite~\cite{olariu2008,kikuchi2005,rule2011,jeschke2011,willenberg2016,inosov2018}. Through this combined effort a new level of insight into complex topics of quantum magnetism is achieved.

The $\Delta$-, or sawtooth chain represents one of the fundamental models of frustrated quantum magnetism. It consists of a chain of spin triangles, with the Hamiltonian 
\begin{equation}
\mathcal{H}=\sum_{i}J\mathbf{S}_{i}\cdot\mathbf{S}_{i+2}+J^{\prime
}\big(\mathbf{S}_{i}\cdot\mathbf{S}_{i+1}+\mathbf{S}_{i+1}\cdot\mathbf{S}_{i+2}\big)-\mathbf{h}\cdot\mathbf{S}_{i}. \label{eq:delta}
\end{equation}
$\mathbf{S}_{i}$ represents a spin $S = 1/2$ at site $i$; the sites $i$ and $i+2$ are neighbors in the chain ``spine'', while $J^{\prime}$ is the interaction between spine sites and the sawteeth tips. \textbf{h} is the external magnetic field. This model has been studied theoretically for decades~\cite{hamada1988,kubo1993,otsuka1995,nakamura1995,nakamura1996,sen1996,maisinger1998,blundell2003,zhitomirsky2004,tonegawa2004,derzhko2004,chandra2004,richter2004,inagaki2005,richter2008,hida2008,hao2011,krivnov2014,dmitriev2015,dmitriev2016,dmitriev2018}. Real materials, however, are inevitably more complex than this simplified model. Delafossite and euchroite have more than two relevant couplings~\cite{bacq2005,kikuchi2011}, certain metalorganic systems have a ferromagnetic intra-spine $J$~\cite{inagaki2005,schnack2018}, and in Rb$_{2}$NaTi$_{3}$F$_{12}$ the $\Delta$-chain is coupled to an antiferromagnetic (AFM) chain~\cite{Jeschke2019}. In this Letter, based on a combined experimental and theoretical study of atacamite, Cu$_{2}$Cl(OH)$_{3}$, we show that its magnetic behavior originate from an intricate and rather unusual 3D connectivity of $\Delta$-chains, not previously addressed.

\begin{figure*}[t]
\includegraphics[width=\textwidth]{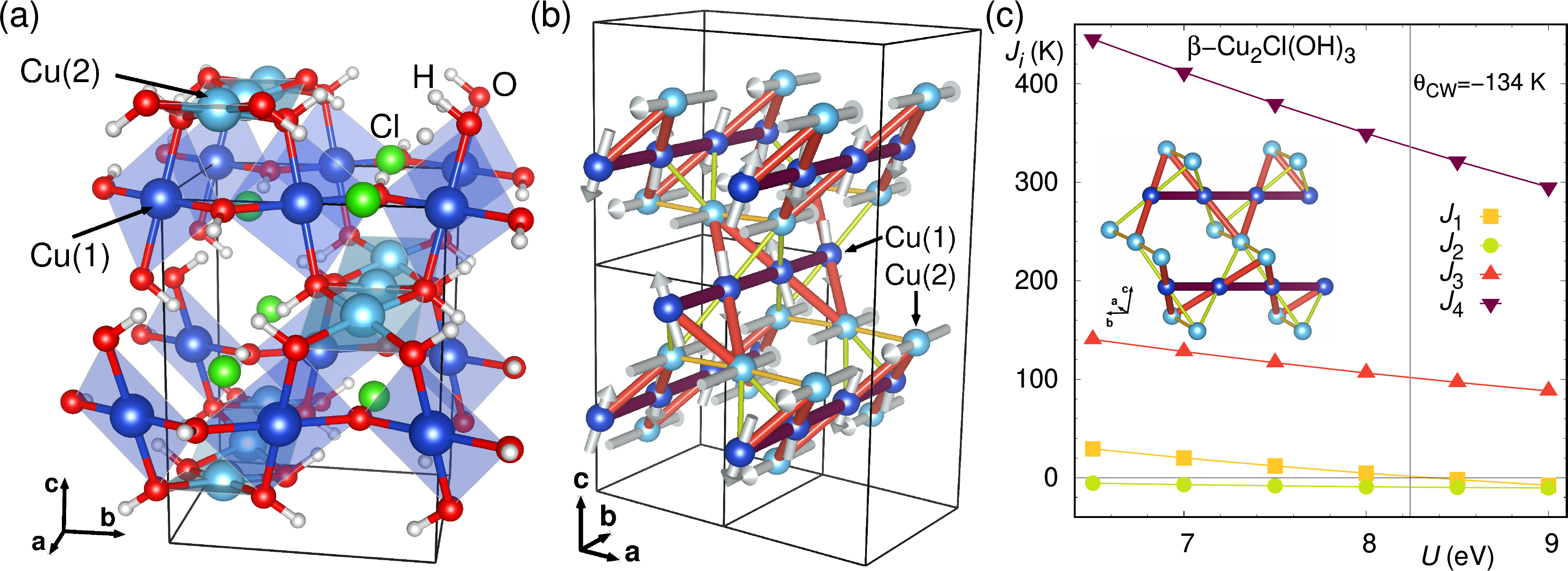}\caption{(color online) (a) Crystal structure of atacamite Cu$_{2}$Cl(OH)$_{3}$. (b) Visualization of the dominant magnetic exchange paths in atacamite forming a sawtooth pattern. Also shown is the magnetic structure together with the nuclear and the magnetic unit cell (black solid lines). (c) Cu-Cu exchange couplings of atacamite, $J_{i}$, with $i = 1, 2, \ldots$ denoting first, second, $\ldots$ neighbor. The vertical line indicates the $U$ value at which the couplings $J_{1} = 1.3$\,K, $J_{2} = -9.6$\,K, $J_{3} = 102$\,K, $J_{4} = 336$\,K, $J_{11b} = 15.6$\,K, $J_{13} = 1.1$\,K match the experimental Curie-Weiss temperature; for details see text.}
\label{fig:structure}
\end{figure*}

We have measured magnetization, magnetic susceptibility and specific heat of atacamite in fields up to 13\,T. Neutron scattering was carried out at the HZB BER~II reactor using the instruments E2, E5 and HFM/EXED for fields up to 25\,T~\cite{E2paper,heinze2018,HFMEXEDpaper}. We have determined both the magnetic and crystallographic structures of our mineral single crystals~\cite{SM,xtal,sears1995,brown1995}. In addition, we have performed a high-field magnetostriction and magnetization study in fields up to 65\,T at the Pulsed Field Facility of the NHMFL, Los Alamos. In the present work, we focus on data taken in magnetic fields applied along the crystallographic $b$ axis. 

Atacamite magnetically orders at low temperatures and we found a complex field-induced spin reorientation behavior. Magnetic fields of $\sim 30$\,T suppress the ordered state, taking the system to nearly half its saturation magnetization, where it persists up to the highest field reached in this study. To rationalize these results, we have investigated the electronic structure and magnetic interactions using Density Functional Theory (DFT) with full potential local orbital (FPLO) basis~\cite{koepernik1999} and generalized gradient approximation (GGA) functional~\cite{perdew1996}; electronic correlations on Cu$^{2+}$ were accounted for by the GGA+$U$ method~\cite{Liechtenstein1995}. The Hamiltonian thus obtained consists of strongly coupled Cu $\Delta$-chains, forming a weakly coupled network. We consider the uncoupled chains in a magnetic field using infinite system time-evolving block decimation (iTEBD)~\cite{vidal2007} as well as exact diagonalization (ED). The results justify our subsequent evaluation of the magnetization process within a 3D mean-field approximation (MFA), and accounting for the inter-chain coupling.

Atacamite Cu$_{2}$Cl(OH)$_{3}$ crystallizes in a $Pnma$ orthorhombic structure (lattice constants $a = 6.02797$\,\AA , $b = 6.86383$\,\AA , $c = 9.11562$\,\AA; Fig.~\ref{fig:structure}~(a))~\cite{SM,parise1986,zheng2005}. There are two inequivalent Cu sites (dark (Cu(1)) and light (Cu(2)) blue spheres). Previously, this crystal structure was derived from a network of pyrochlore tetrahedra built up by Cu$^{2+}$ ions~\cite{zheng2004,zheng2005,zenmyo2013}. Our DFT calculations, however, indicate that the symmetry of the magnetic Hamiltonian is dramatically lower than the one anticipated from the bond lengths only. Indeed, the bonds derived from the 1$^{\mathrm{st}}$, 2$^{\mathrm{nd}}$ and 3$^{\mathrm{rd}}$ pyrochlore coordination shells vary in length by $\pm 10\%$ within each set, but the calculated exchange parameters (corresponding to 4, 6 and 7 distinct Cu-Cu distances, respectively), vary by two orders of magnitude. As we show below, our calculated Hamiltonian provides an excellent explanation of the experimental observations.

\begin{figure}[t]
\centering
\includegraphics[width=\linewidth]{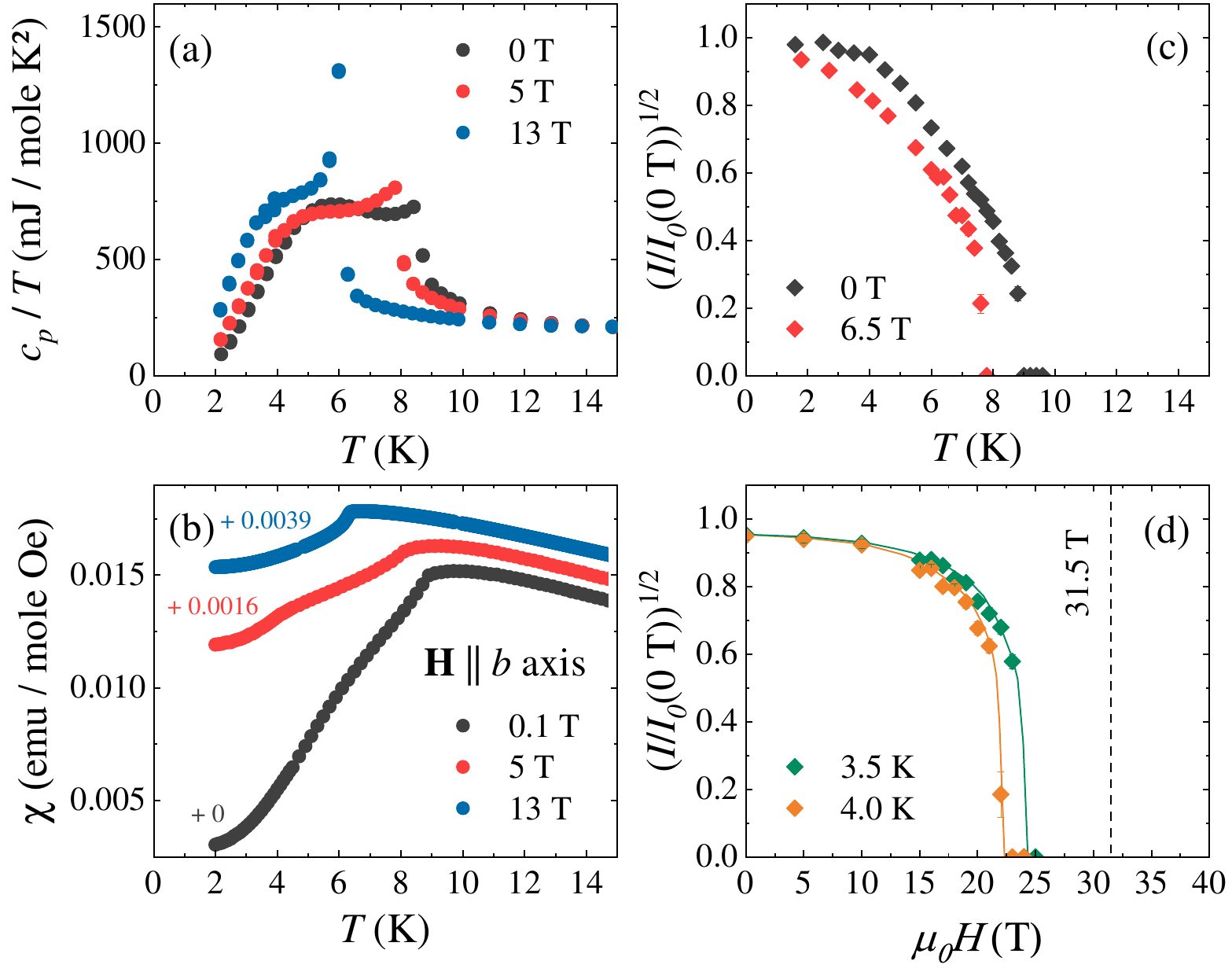}\caption{(color online) (a) Specific heat $c_{p}/T$, (b) magnetic susceptibility $\chi$, and normalized intensity of the $(1/2\;0\;1/2)_{\mathrm{M}}$ neutron scattering reflection as function of (c) temperature and (d) magnetic field ($\mathbf{H} \parallel b$ axis). Data in (b) are shifted for clarity by values denoted in the plot; lines in (d) are guides to the eye, dashed line indicates the plateau field.}
\label{fig:overview}
\end{figure}

First evidence of the existence of ordered magnetism in atacamite was reported previously~\cite{heinze2018,zheng2004,zheng2005,zenmyo2013}. Furthermore, we present zero-field specific heat measurements, with an anomaly indicating a magnetic transition at $T_{\mathrm{N}} = 8.4$\,K (Fig.~\ref{fig:overview}~(a)). An antiferromagnetic anomaly is also observed at $T_{\mathrm{N}} = 8.4$\,K in the low-field (0.1\,T) susceptibility as maximum in $d(\chi T)/dT$ (Fig.~\ref{fig:overview}~(b))~\cite{impurity}. In neutron diffraction, we find magnetic intensity below a slightly higher $T_{\mathrm{N}} = 8.9$\,K with a magnetic propagation vector $\mathbf{q} = (1/2, 0, 1/2)$ (Fig.~\ref{fig:overview}~(c)).

We also detect an additional hump in the specific heat at $T \sim 5$\,K (Fig.~\ref{fig:overview}~(a)) hinting at a more complex temperature evolution of the magnetic state, involving, for instance, spin reorientations. A calculation of the magnetic entropy from our data (ignoring a phonon contribution) gives a value $\sim 0.65\,R\,\ln(2)$ at $T_{\mathrm{N}}$~\cite{SM}. Such a small value is typical for magnetically ordered states in frustrated magnets with the magnetic entropy being distributed over the temperature scale set by the dominant coupling strengths, here $J_{4}$ and $J_{3}$ (Fig.~\ref{fig:structure}~(c)).

In magnetic fields $\mathbf{H} \parallel b$, the features in the specific heat and the susceptibility are shifted to lower temperatures and the AFM anomaly is sharpened (Fig.~\ref{fig:overview}~(a)--(b)). This shift is supported by neutron scattering in 6.5\,T (Fig.~\ref{fig:overview}~(c)). Field-dependent neutron scattering at the HFM/EXED instrument yields a suppression of AFM order at 24\,T ($T = 3.5$\,K) (Fig.~\ref{fig:overview}~(d)). Altogether, an external magnetic field leads to a suppression of the AFM phase, with $T_{\mathrm{N}}$ fully suppressed in $\sim 30$\,T.

The low-$T$ susceptibility in high fields is larger than in low fields (Fig.~\ref{fig:overview}~(b)). It reflects a metamagnetic transition occurring at a few Tesla (see below). Since for the other crystallographic directions we find no such transition, it suggests that the $b$ axis is the easy magnetic axis and this is a spin-flop transition~\cite{SM}. This is consistent with our refined magnetic structure.

From the magnetic Bragg peak intensities, we derive the magnetic structure in Fig.~\ref{fig:structure}~(b)~\cite{SM,bertaut1986}. On the Cu(1) site, the ordered magnetic moments of $0.34(4)\,\mu_{\mathrm{B}}$ are arranged in a nearly perfect AFM pattern with the Cartesian components $\mu_{\mathrm{ord,Cu(1)}}(x,y,z) = \left[0.09(9),0.04(2),0.32(7)\right]\,\mu_{\mathrm{B}}$. The ordering vector corresponds to alternating signs for the $x$ and $z$ moment components, while the $y$ component stays the same within the same chain. The angle between two Cu(1) neighbors is thus $\theta = 166.3^{\circ}$, close to $180^{\circ}$. All Cu(2) sites carry a moment $\mu_{\mathrm{ord,Cu(2)}}(x,y,z) = \left[0,0.59(2),0\right]\,\mu_{\mathrm{B}}$ where moments are parallel to $y$ within one set of sawtooth sites of a single chain (details in Ref.~\cite{SM}).

\begin{figure}[t]
\includegraphics[width=\linewidth]{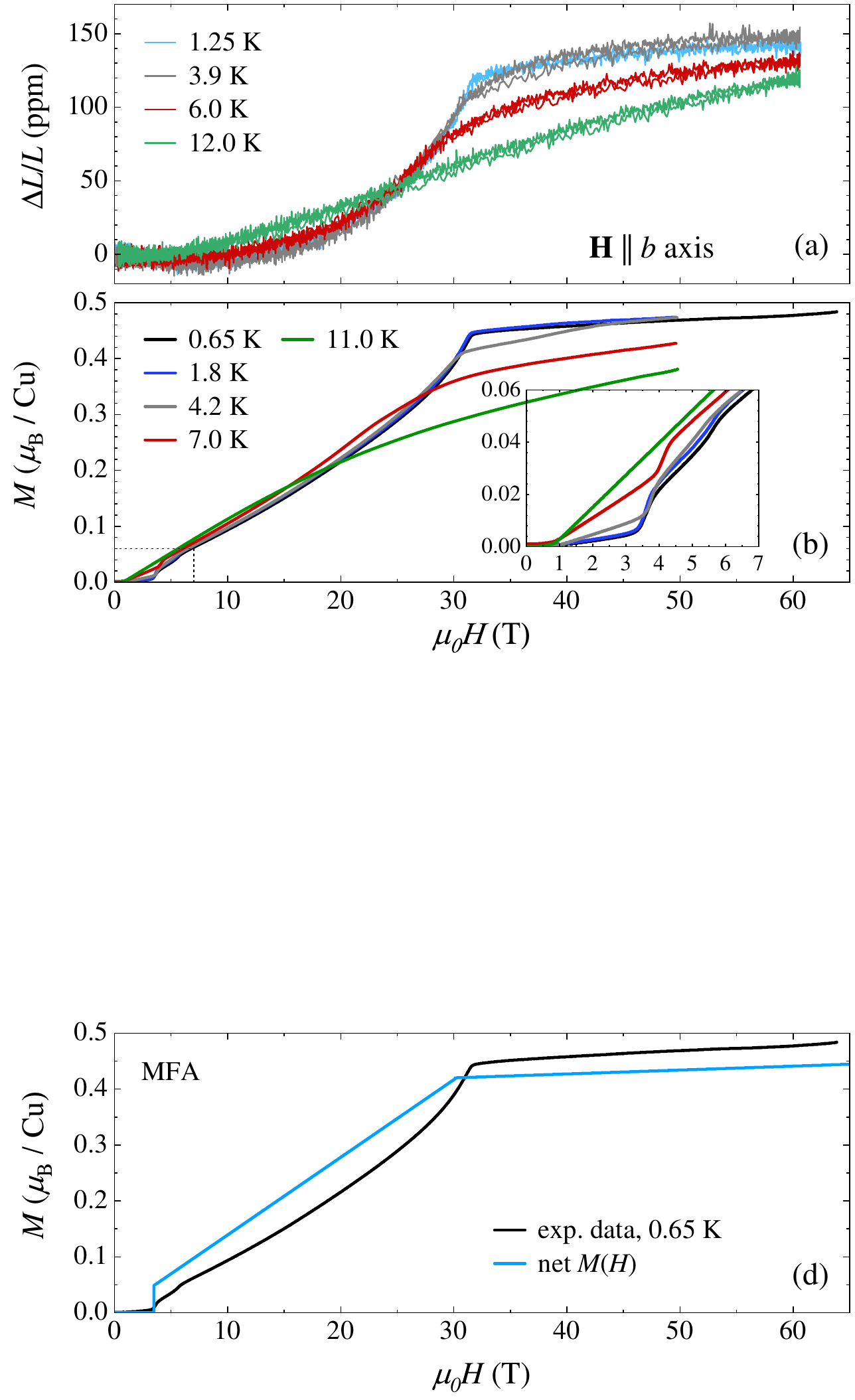}\llap{\raisebox{3.84cm}{\includegraphics[width=\linewidth]{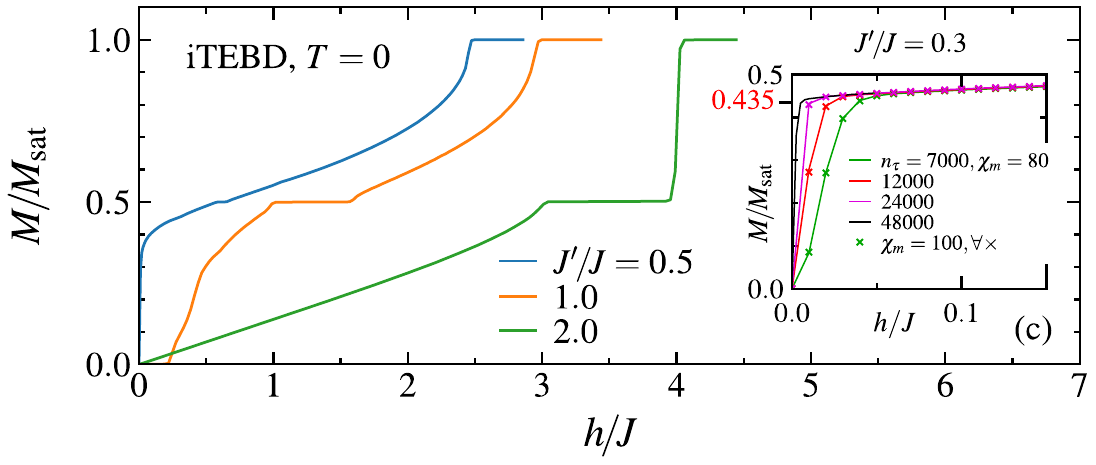}\hspace*{-.8mm}}}
\caption{(color online) (a) Magnetostriction and (b) magnetization of atacamite in magnetic fields $\mathbf{H} \parallel b$ axis at different temperatures. (c) Magnetization versus $h$ via iTEBD for the $\Delta$-chain for different $J^{\prime}/J$ at $T = 0$. Note that for $J \sim 336$\,K, a unit of $h$ corresponds to $\sim 250$\,T. Inset: Magnetization at $J^{\prime}/J = 0.3$ for max.~Schmidt rank $\chi_{m} = 80~(100)$ lines (crosses) and imaginary time $\tau = n_{\tau}d\tau$, $d\tau = 0.01/J$. (d) MFA for 3D coupled $\Delta$-chains.}
\label{fig:highfield}
\end{figure}

To assess the magnetic phase diagram, we used magnetometry in pulsed magnetic fields for $\mathbf{H} \parallel b$~\cite{jaime2017,detwiler2000}. In Fig.~\ref{fig:highfield}~(a)--(b) we summarize the magnetostriction and magnetization, respectively. Below $T_{\mathrm{N}}$ and fields of $\mu_{0}H_{1} \lesssim 4$\,T, a kink in the magnetization indicates a spin-flop transition. When increasing temperature the kink becomes weaker and shifts to higher fields in the AFM phase~\cite{SM}. For temperatures below $\sim 5$\,K, a weak shoulder appears (inset Fig.~\ref{fig:highfield}~(b)), which corresponds to shallow minima in the magnetostriction~\cite{SM}. This might indicate a splitting of the spin-flop transition due to a weak three-axes exchange anisotropy.

Immediately after the spin-flop transition, $M(H)$ grows linearly with $dM/dH \approx 0.013\,\mu_{\mathrm{B}}/$T, but starts bending upwards up to a field of $\mu_{0}H_{2} = 31.5$\,T, where the slope reaches $0.042\,\mu_{\mathrm{B}}/$T (Fig.~\ref{fig:highfield}~(b)). After that, a wide magnetization plateau-like behavior at about $0.45\,\mu_{\mathrm{B}}$/Cu sets in. The plateau, also detected in the magnetostriction (Fig.~\ref{fig:highfield}~(a)), reaches up to highest measured fields and is not perfectly flat, but rising at a rate of $0.001\,\mu_{\mathrm{B}}/$T.

\begin{figure}[t]
\centering
\includegraphics[width=\linewidth]{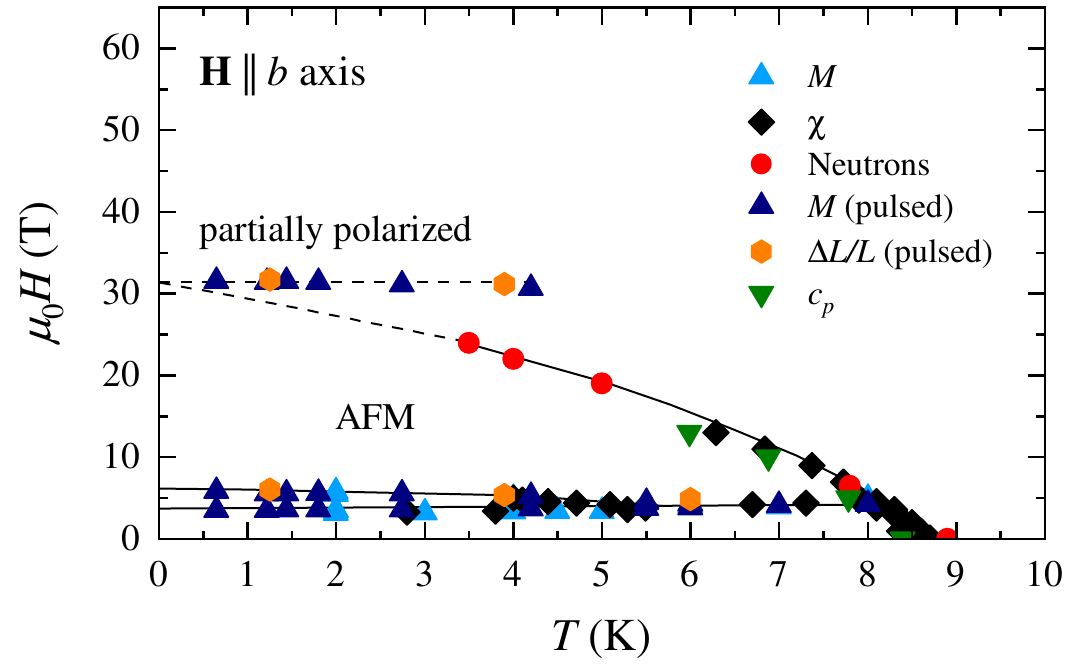}
\caption{(color online) Magnetic phase diagram of atacamite for $\mathbf{H} \parallel b$; for details see text. Data points at 31.5\,T from pulsed-field experiments as measured, not corrected for MCE~\cite{MCE}.}
\label{fig:phase}
\end{figure}

From our data, we construct the magnetic phase diagram of atacamite for $\mathbf{H} \parallel b$ (Fig.~\ref{fig:phase}). The AFM phase exists below $T_{\mathrm{N}}$ and up to $\sim 30$\,T. It is separated into a low-field regime with the magnetic structure described before and a high-field regime for fields above the spin-flop transition. In the limit $T \to 0$\,K, the suppression of AFM order possibly coincides with the appearance of a magnetization plateau-like behavior. To fully establish the magnetic phase diagram in this field region, it requires a determination of the magnetocaloric effect in pulsed magnetic fields~\cite{nomura2020,MCE}. At highest fields of 65\,T the system is still far from saturation.

It is now instructive to establish the Hamiltonian of atacamite and connect it to the observations. To this end, we used an energy mapping technique~\cite{Glasbrenner2015,Iqbal2018,Ghosh2019} to calculate 17 exchange interactions, derived from the first three coordination shells of the parent pyrochlore structure~\cite{SM}. Only six of them exceed 1\,K, or 0.3\% of the dominant coupling $J_{4}$ (Fig.~\ref{fig:structure}~(c)). A Curie-Weiss temperature $\Theta_{\mathrm{CW}}$ calculated from our $J$s for $U = 8.24$\,eV, typical for Cu$^{2+}$, matches the experiment~\cite{heinze2018,zheng2005}. Two antiferromagnetic interactions stand out: $J_{4} = 336$\,K and $J_{3} = 102$\,K, which bind Cu(1) and Cu(2) atoms into anisotropic $\Delta$-chains (compare Eq.~(\ref{eq:delta}) with $J \equiv J_{4}$ and $J^{\prime} \equiv J_{3}$).

Based on these findings, we consider a single $\Delta$-chain (Eq.~(\ref{eq:delta})). Fig.~\ref{fig:highfield}~(c) shows iTEBD results, complementary ED results are described in \cite{SM}. For $0.5\leq J^{\prime}/J\leq2$ we observe the famous quantum half-magnetization plateau~\cite{richter2004,richter2008,Metavitsiadis2020}. However, it is practically invisible for $J^{\prime}/J\lesssim0.5$, relevant for atacamite. Moreover, its field scale is of $O(J) \sim O(250$\,T$)$. Therefore, while tempting, the observed flattening of $M(H)$ around 31.5\,T in atacamite is \textit{not} related to the half-magnetization plateau physics. On the other hand, the 3D exchange among the chains proves to be relevant. The second finding in Fig.~\ref{fig:highfield}~(c) is far more striking and has not been appreciated before: in the small-$J^{\prime}$ gapless phase, \textit{e.g.}, at $J^{\prime}/J=0.5$, the low-$h$ susceptibility appears singular and the magnetization approaches a \textit{non}-quantized \textit{finite} value as $h\rightarrow0$. For the relevant $J^{\prime}/J\sim0.3$, the inset of Fig.~\ref{fig:highfield}~(c) shows iTEBD versus increasing imaginary simulation times $\tau=n_{\tau}d\tau$ in terms of $d\tau=0.01/J$. Since iTEBD inherits the limit of system size $N\rightarrow\infty$ by construction, and by identifying $\tau^{-1}\sim T$ with a quasi-temperature, we extract the following order of limits $\lim_{h\rightarrow0}\lim_{N\rightarrow\infty}\langle S^{z}\rangle_{T}:=M_{0}(T)$ from the inset. For $T\neq0$, we find $M_{0}(T)=0$, however as $T \rightarrow 0$, very likely, $M_{0}(T=0)/\mathrm{cell}\approx0.435\neq0$, all of which is consistent with Mermin-Wagner's theorem. Rephrasing, we seem to observe ferromagnetic order at $T=0$ for the $\Delta$-chain at $J^{\prime}/J\sim0.3$. This likely holds for the entire small-$J^{\prime}$ gapless phase. This is consistent with ED~\cite{SM} and with a classical treatment of the $\Delta$-chain.

In the MFA, the ground state of a single $\Delta$-chain has the same pattern as observed experimentally, with $\theta=360^{\circ}-2\arccos\left(-\frac{J^{\prime}}{2J}\right)=162.5^{\circ}$, in excellent agreement with our neutron data, with the Cu$^{2+}$ net moment, $M_{\mathrm{eff}}=(M_{2}-M_{1}J^{\prime}/2J)/2$~\cite{SM}. If the moment of each copper is taken to be $M_{2}=M_{1}=1\,\mu_{\mathrm{B}}$, then $M_{\mathrm{eff}}^{\mathrm{MFA}}=0.42\,\mu_{\mathrm{B}}$. We know from experiment though that these moments are suppressed by fluctuations to $M_{1}=0.34\,\mu_{\mathrm{B}}$, $M_{2}=0.59\,\mu_{\mathrm{B}}$. This reduces the net moment to $M_{\mathrm{eff}}(H=0)=0.27\,\mu_{\mathrm{B}}$/Cu~\cite{SM}. On the other hand, $M_{\mathrm{eff}}^{\mathrm{MFA}}=0.42\,\mu_{\mathrm{B}}$ agrees very well with the magnetization at $\sim30$\,T, indicating that in such fields the fluctuations are mostly quenched. In the following we used $M_{\mathrm{eff}}(H)$ and $M_{1,2}(H)$ linearly interpolating between the two limits.

We are now in a position to describe an effective 3D magnetic model that can be addressed by classical mean-field calculations. These treat the $\Delta$-chains as emergent, rigid macroscopic objects, carrying a large magnetic moment. Classically, the latter arises primarily from Cu(2) moments being aligned ferromagnetically along $b$ (Fig.~\ref{fig:structure}~(b)). These large moments are AFM stacked into a 2D crystal and coupled via the small subleading exchange interactions. Their projection onto the $ac$ plane forms an anisotropic triangular lattice with three effective AFM couplings, $J_{\rm B}\lesssim J_{\rm A}\ll J_{\rm C}$~\cite{SM}. For our selected value of $U=8.24$\,eV, $J_{\rm C}\approx8$\,K, $J_{\rm A}\approx0.5$\,K and $J_{\rm B}\approx0$\,K. The classical ground state of this model is collinear with N\'{e}el order along C and B, and FM order along A, as observed experimentally (Fig.~\ref{fig:structure}~(b)). We note, however, that $J_{\rm A,B}$ rapidly rise with decreasing $U$, and at $U=7$\,eV (still an admissible value for Cu$^{2+})$ $J_{\rm A}\approx J_{\rm B}\approx J_{\rm C}\approx 10$\,K. In that case, the MFA ground state would have been the 120$^\circ$ order.

We focus on the 2D collinear N\'{e}el order along the C direction. Since $b$ is the easy axis, the MFA predicts a spin-flop at low fields $\mathbf{H}\parallel b$, whereupon all magnetic moments rigidly rotate so that the Cu(2) moments are $\perp \mathbf{H}$. The classical spin-flop field is $\mu_{0}H_{1}=\frac{M_{2}(0)}{M_{\mathrm{eff}}(0)}\sqrt{2KJ_{\Delta-\Delta}}$, where $K$ measures the uniaxial anisotropy (given here for simplicity as an effective single-site term), and the effective coupling $J_{\Delta-\Delta} \approx J_{\rm C}+J_{\rm B}\approx8.5$\,K for $U = 8.24$\,eV. To reproduce the experimentally observed $\mu_{0}H_{1} \sim 3.5$\,T one needs $K \sim 0.04$\,K (a typical energy scale for Cu$^{2+}$~\cite{Tranquada1989}). The low symmetry of atacamite allows also for some in-plane magnetic anisotropy. A possible splitting of $H_{1}$ into two close transitions likely reflects such anisotropies.

As the field increases, the spin-flopped state gradually cants, generating a net magnetic moment of $HM_{\mathrm{eff}}^{2}(H)/M_{2}^{2}(H)J_{\Delta-\Delta}$. In an uncorrected MFA, $M(H)$ is linear. However, accounting for quantum fluctuations and their gradual quenching with field leads to deviation from linearity. These deviations are visible in experiment~\cite{SM}. At a field $H_{2}$ the moments cant into the ``plateau'' configuration, where all $\Delta$-chains are ordered ferromagnetically, and the total moment is $M = M_{\mathrm{eff}}^{\mathrm{MFA}}$. For our calculated parameters, $\mu_{0}H_{2} = 30.1$\,T, to be compared to the experimental value of 31.5\,T. In this state, the total moment does not remain constant but keeps rising as $M_{\mathrm{eff}}(H)=(1/2-J^{\prime}/4J+H/4J)\mu_{\mathrm{B}}$. The differential susceptibility $dM/dH$, calculated this way, is much smaller than in the experiment, yet is qualitatively consistent with the latter~\cite{SM}.

The overall dependence of $M(H)$ as calculated in MFA, adjusted for the quenching of the fluctuations, and using the DFT exchange couplings, exhibits an excellent agreement with the experiment (Fig.~\ref{fig:highfield}~(d)), giving credence to the calculation and to the described scenario.

Finally, let us discuss the finite-$T$ phase diagram. Since temperature effects might slightly change the ratios between $J_{\rm A}$, $J_{\rm B}$, $J_{\rm C}$, we note that a tuning towards the region $J_{\rm A} \approx J_{\rm B} \approx J_{\rm C}$ opens the possibility of multiple phases with various degrees of non-collinearity, some of them only emerging at finite temperatures~\cite{starykh2015}. While our current observations do not yield hints as to the specific nature of this phase, we note that the phase diagram for the simple isotropic triangular lattice is similar to our Fig.~\ref{fig:phase} (see Fig.~3 in Ref.~\cite{starykh2015}). This is a subject for future investigations.

We have studied the natural mineral Cu$_{2}$Cl(OH)$_{3}$ and found that it is well described as a weakly coupled asymmetric triangular lattice of $S = 1/2$ $\Delta$-chains. We find an unusual magnetic behavior, with a magnetization deceivingly reminiscent of the quantum half-magnetization plateau, which however turns out to be a classical effect, well described by MFA. A magnetic Hamiltonian derived from first-principles calculations predicts a spin flop, a magnetization plateau, and weak deviation from the plateau behavior in high fields. This compound therefore represents a unique example of strongly-coupled 1D ferromagnetic objects coordinated by weak and anisotropic 2D interactions. We hope that this discovery will encourage more studies of this class of magnetic models.

\begin{acknowledgments}
We gratefully acknowledge the financial support from HZB. This work has partially been supported by the DFG under Contract Nos.~WO1532/3-2 and SU229/9-2. We gratefully acknowledge T. Reimann for fruitful discussions, S. Gerischer, R. Wahle, S. Kempfer, P. Heller and P. Smeibidl for their support at the HFM/EXED at HZB as well as experimental support by G. Bastien in the initial stages of this work. We thank G. Paskalis and J. McAllister for supplying us with two of the atacamite crystals used for this study. W.B. and A.U.B.W. have been supported in part by the DFG through projects A02 and B01 of SFB 1143 (project-id 247310070), respectively. W.B. acknowledges partial support by QUANOMET and hospitality of the PSM, Dresden. The National High Magnetic Field Pulsed Field user facility is supported by the National Science Foundation through cooperative grant DMR 1157490, the State of Florida, and the US Department of Energy. V.S.Z. was supported by the Laboratory-Directed Research and Development program at Los Alamos National Laboratory. S.S. acknowledges support by the Magnet Lab.~Visiting Scientist Program of the NHMFL. I.I.M. acknowledges support by the Research Institute for Interdisciplinary Science through the Okayama University visiting scientist program and from DOE through the grant DE-SC0021089.
\end{acknowledgments}


\begin{thebibliography}{99}
\bibitem{impurity} From an analysis of the low temperature magnetic susceptibility of several atacamite samples we estimate that the amount of $S = 1/2$ impurities is less than 0.5\,\% as indicated by the absence of a Curie tail down to the lowest experimental temperature of $T = 2$\,K (see Fig.~\ref{fig:overview}~(b)).
\bibitem{MCE} Since the original experiments were completed we became aware of the existence of a strong magnetocaloric effect (MCE) in atacamite that might cause the sample to be adiabatically (as opposed to isotermically) magnetized in pulsed magnetic fields, even when immersed in liquid He at cryogenic temperatures. A detailed MCE study, outside of the scope of the present manuscript, is currently under way to establish the $(T, \mu_0 H)$ phase diagram of this material beyond ambiguities.
\end{thebibliography}
\end{document}